\documentclass[prd,amsmath,amssymb,preprintnumbers,superscriptaddress,twocolumn,10pt,nofootinbib,showkeys,showpacs]{revtex4}

\pdfoutput=1

\usepackage{graphicx}
\usepackage{dcolumn}
\usepackage{bm}
\usepackage{amssymb}
\usepackage{latexsym}
\usepackage{booktabs}
\usepackage{amsmath}
\usepackage{multirow}
\usepackage{url}

\usepackage{float}
\usepackage[colorlinks=true, linkcolor=red, citecolor=blue]{hyperref}

\usepackage[normalem]{ulem}
\usepackage{color}
\usepackage{array}
\usepackage{enumerate}
\begin{document}

\title{Cosmology with fast radio bursts in the era of SKA}


\author{Ji-Guo Zhang}
\affiliation{Department of Physics, College of Sciences, Northeastern University, Shenyang 110819, China}
\author{Ze-Wei Zhao}
\affiliation{Department of Physics, College of Sciences, Northeastern University, Shenyang 110819, China}
\author{Yichao Li}
\affiliation{Department of Physics, College of Sciences, Northeastern University, Shenyang 110819, China}
\affiliation{Key Laboratory of Cosmology and Astrophysics (Liaoning), Northeastern University, Shenyang 110819, China}
\author{Jing-Fei Zhang}
\affiliation{Department of Physics, College of Sciences, Northeastern University, Shenyang 110819, China}
\affiliation{Key Laboratory of Cosmology and Astrophysics (Liaoning), Northeastern University, Shenyang 110819, China}
\author{Di Li}
\affiliation{National Astronomical Observatories, Chinese Academy of Sciences, Beijing 100101, China}
\affiliation{University of Chinese Academy of Sciences, Beijing 100049, China}
\affiliation{Research Center for Intelligent Computing Platforms, Zhejiang Laboratory, Hangzhou 311100, China}
\author{Xin Zhang}\thanks{Corresponding author.\\zhangxin@mail.neu.edu.cn}
\affiliation{Department of Physics, College of Sciences, Northeastern University, Shenyang 110819, China}
\affiliation{Key Laboratory of Cosmology and Astrophysics (Liaoning), Northeastern University, Shenyang 110819, China}
\affiliation{National Frontiers Science Center for Industrial Intelligence and Systems Optimization, Northeastern University, Shenyang 110819, China}
\affiliation{Key Laboratory of Data Analytics and Optimization for Smart Industry (Ministry of Education), Northeastern University, Shenyang 110819, China}

\begin{abstract}
We present a forecast of the cosmological parameter estimation using fast radio bursts (FRBs) from the upcoming Square Kilometre Array (SKA), focusing on the issues of dark energy, the Hubble constant, and baryon density. We simulate $10^5$ and $10^6$ localized FRBs from a 10-year SKA observation, and find that: (i) using $10^6$ FRB data alone can tightly constrain dark-energy equation of state parameters better than CMB+BAO+SNe, providing an independent cosmological probe to explore dark energy; (ii) combining the FRB data with gravitational-wave standard siren data from 10-year observation with the Einstein Telescope, the Hubble constant can be constrained to a sub-percent level, serving as a powerful low-redshift probe; (iii) using $10^6$ FRB data can constrain the baryon density $\Omega_{\rm b}h$ to a precision of $\sim 0.1\%$. Our results indicate that SKA-era FRBs will provide precise cosmological measurements to shed light on both dark energy and the missing baryon problem, and help resolve the Hubble tension.

\pacs{98.80.Es, 95.36.+x, 95.55.Jz}
\keywords{fast radio burst, cosmology, dark energy, the Hubble constant, baryon density, Square Kilometre Array}

\end{abstract}

\maketitle
\section{Introduction}\label{sec1}
Fast radio bursts (FRBs) are millisecond-duration bright radio transients with dispersion measures (DMs) beyond Galactic expectations \citep{lorimer2007,thornton2013pop}, which remain enigmatic due to their unknown origins and underlying emission mechanisms. Nonetheless, with the high all-sky FRB event rate and advancements of radio astronomy instrument, many fascinating FRB discoveries have been achieved, such as repeaters \citep{spitler2016repeating,CHIMEFRB2019ghr,Fonseca2020cdd,Niu2021bnl,Marcote:2020ljw,Kirsten:2021llv,Feng:2023qym}, some of which have periodic activity \citep{rajwade2020possible,CHIMEFRB2020bcn,pastor2021chromatic}, polarization characteristics \citep{wang2022rep,Zhang:2023eui} and in particular, the landmark detection of FRB 200428 associated with a Galactic magnetar SGR 1935+2154 \citep{CHIMEFRB:2020abu,bochenek2020fast,lin2020no,li2021hxmt}, indicating that at least some FRBs can originate from magnetars. Thus far, more than 700 FRBs have been observed (including 29 repeaters) \citep{Hu:2022oie} largely thanks to the splendid release of the first catalogue \citep{CHIMEFRB:2021srp} from the Canadian Hydrogen Intensity Mapping Experiment Fast Radio Burst project (CHIME/FRB) \citep{bandura2014canadian}. More than a dozen of FRBs have been located within individual galaxies \citep{Scholz:2016rpt,Chatterjee:2017dqg,Tendulkar:2017vuq,Marcote:2017wan,prochaska2019low,Bannister:2019iju,Ravi:2019alc,Marcote:2020ljw,Bhandari:2020oyb,heintz2020host,Law:2020cnm,Nimmo:2021ntn,Bhandari:2021pvj,Bhardwaj:2021xaa,Ryder:2022qpg,Kirsten:2021llv,Jankowski:2023sot,Driessen:2023lxj}. These facts show a growing trend that FRBs can serve as a powerful cosmological probe.

By employing FRBs, one can establish a relation between dispersion measure and redshift, which is like the distance--redshift relation, because in this relation DM serves as a proxy of distance \citep{gao2014fast}. The observed DM encodes the content of free electrons along the line of sight. The dominant part in DM comes from the contribution of the intergalactic medium (IGM), denoted as $\rm DM_{IGM}$, and this quantity is related to cosmology \citep{Deng:2013aga,Zhou:2014yta,Yang2016zbm}. If FRBs can be localized, i.e., their host galaxies can be found and the redshifts of them can be measured, then the localized FRBs can be used as a cosmological probe through the Macquart relation (i.e., the $\rm DM_{IGM}$--$z$ relation) \citep{macquart2020census}. Using localized FRBs, substantial cosmological studies have been done, including probing the baryon fraction of the IGM \citep{li:2019cos,Wei:2019uhh,Qiang:2020vta,Dai:2021czy,Wang:2022ami}, strong gravitational lensing \citep{Munoz:2016tmg,li2018strongly,Liu:2019jka,Liao:2020wae,Zhao:2021jeb,Gao:2022ifq}, testing the equivalence principle \citep{Wei:2015hwd,Hashimoto:2021swe,Reischke:2023gjv}, probing the reionization history \citep{beniamini2021exploring,Heimersheim:2021meu} and other cosmological applications \citep{Shao:2017tuu,Wu:2020jmx,Jing:2022typ,Zhu:2022mzv,Guo:2023hgb}. For recent reviews, see Refs.~\citep{Bhandari:2021thi,xiao2021om,Petroff:2021wug,Caleb:2021xqe,Bailes:2022pxa,Zhang:2022uzl}. The Square Kilometre Array (SKA), as the upcoming largest high-sensitivity radio telescope \citep{macquart2015fast}, promises detection of FRBs several orders of magnitude larger than the current sample size in the next decades \citep{Fialkov:2017qoz,hashimoto2020fast}. Obviously, it is expected that some important issues in cosmology would be solved with the help of the plentiful FRBs in the era of SKA.

Dark energy is one of the most difficult theoretical issues in fundamental physics and cosmology. To elucidate the nature of dark energy, 
the most crucial step is to precisely measure the equation of state (EoS) of dark energy, which can be achieved by precisely measuring the expansion history of the universe. The $\Lambda$ cold dark matter ($\Lambda$CDM) model has long been viewed as the standard model of cosmology, in which the cosmological constant $\Lambda$ with the EoS $w=-1$ serves as dark energy  \citep{bahcall1999cosmic}. The precise measurements on anisotropies in the cosmic microwave background (CMB) by the Planck satellite strongly favor the six-parameter $\Lambda$CDM cosmology \citep{Planck:2018vyg}. However, the CMB is an early-universe probe, and so when a dynamical dark energy (with EoS unequal to $-1$ and usually evolving) is considered, the extra parameters describing EoS cannot be effectively constrained by CMB \citep{Planck:2018nkj}. Therefore, developing independent late-universe cosmological probes becomes rather important for exploring the nature of dark energy. Usually, the late-universe observations such as baryon acoustic oscillations (BAO) and type Ia supernovae (SNe) themselves cannot solely tightly constrain dark energy, but they serve as complementary tools to help break the cosmological parameter degeneracies inherent in the CMB \citep{Planck:2018vyg}. Nevertheless, the cosmological tensions between the early and late universe appeared in recent years, such as the so-called ``Hubble tension'' \citep{Riess:2020fzl,Verde:2019ivm}. {In addition to searching for new physics \citep{Yang:2018euj,Guo:2018ans,Feng:2019jqa,Liu:2019awo,vag2020new,Gao:2021xnk,Cai:2021wgv,Wang:2021kxc}, forging late-universe probes that can precisely measure cosmological parameters (in particular related to dark energy) is extremely important for current cosmology \citep{Zhang:2019cww,Cai:2021weh,Moresco:2022phi}}.

In addition to the ongoing and upcoming large optical surveys, other-type late-universe observations may prove to be promising to explore dark energy. For example, the recent intensive discussions on gravitational-wave (GW) standard sirens \citep{Schutz:1986gp,Holz:2005df} show that GWs as a new messenger may offer a novel useful tool for probing the expansion history of the universe \citep{Cai:2017buj,LIGO:2017adf,Chen:2017rfc,Zhao:2018gwk,Chang:2019xcb,Zhang:2019ylr,Zhang:2019loq,Song:2022siz,Jin:2022qnj,Jin:2023zhi} (see Refs.~\citep{Zhang:2019ylr,Bian:2021ini} for recent reviews). In addition, taking neutral hydrogen as a new tracer of the total matter in the universe through observing their 21-cm emissions (in particular, by the intensity mapping method \citep{Chang:2010jp,Bull:2014rha, Xu:2020uws,Zhang:2021yof,Wu:2022jkf,Zhang:2023gaz,Li:2023zer}) to map the three-dimensional distribution of the matter is hopeful to play a crucial role in studying cosmology.

In addition to these, it has been shown in Refs.~\citep{Walters:2017afr,Jaroszynski:2018vgh,zhao2020cosmological,qiu2022forecast,Wu:2022dgy,Zhao:2022bpd} that a sufficient large sample of FRBs (localized ones with the redshifts being measured) could be used as a useful tool to constrain dark-energy parameters. Particularly, Ref.~\citep{zhao2020cosmological} pointed out that, in order to become a useful cosmological tool, at least $\sim 10^4$ localized FRBs are needed, which can be used to help break cosmological parameter degeneracies led by CMB. Moreover, Ref.~\citep{zhao2020cosmological} concluded that with $\sim 10^4$ localized FRBs the results of CMB+FRB are better than those of CMB+BAO; GW+FRB can provide an effective late-universe probe that is comparable with CMB+BAO; CMB+GW+FRB can provide tight constraints on cosmological parameters. The sample size of $\sim 10^4$ localized FRBs is consistent with the detection event rate given by different surveys, such as Australian Square Kilometre Array Pathfinder (ASKAP) \citep{bannister2017detection} and CHIME outrigger \citep{mena2022clock}. However, the in-construction SKA will definitely have much more powerful capability of detecting and localizing FRBs, which will undoubtedly provide a much larger FRB sample to be used in exploring the nature of dark energy. Therefore, the questions of what extent the nature of dark energy can be explored using the FRB observation in the era of SKA and whether the SKA-era FRB observation can be forged into a precise late-universe probe cry out for answers. Our present work wishes to investigate these issues in depth and in detail and to try to give answers to these questions. 

An issue closely related to dark energy is the measurement of the Hubble constant using FRBs, which has been extensively investigated in recent literature \citep{Hagstotz:2021jzu,Wu:2021jyk,James:2022dcx,Liu:2022bmn,Zhao:2022yiv}. The Planck CMB data can tightly constrain the Hubble constant $H_0$ in the $\Lambda$CDM cosmology (at a $\sim 0.8\%$ precision), but when a dynamical dark energy is considered in a cosmological model, the CMB data cannot provide a tight constraint on $H_0$ since the EoS of dark energy is in significant anti-correlation with the Hubble constant. This point has often been employed in the discussions of the possible solution to the ``Hubble tension''. In Ref.~\citep{zhao2020cosmological}, it was shown that FRBs cannot provide an effective constraint on $H_0$ either, but fortunately the orientations of degeneracies related to $H_0$ in the cases of FRBs and CMB are different, leading to the result of the degeneracy inherent in CMB being broken by FRBs. Hence, the combination of FRBs and CMB can give a rather tight constraint on the Hubble constant, which is useful in the case of not considering the Hubble tension. Alternatively, the combination of GW standard sirens and FRBs can be considered to investigate the estimation solely from the late universe, potentially providing a cross-check to the Hubble tension.

In addition, an important topic in the area of FRB cosmology is the ``missing baryon'' problem. Relative to the global density inferred from the big bang nucleosynthesis (BBN) and CMB measurements \citep{fukugita1998cosmic,cen2006baryons}, the detection of baryons in local universe shows a deficit of about $30\%$ \citep{shull2012baryon}. To search for the missing baryons and account for the discrepancy between observation and theoretical prediction, one has recognized that the great majority of baryons reside in the IGM; especially the warm and hot ionized medium (WHIM) rather than galaxies \citep{bregman2007search,Bregman:2023bce}, making the detection quite challenging because of the diffuse resident. \citet{macquart2020census} derived a complete estimate with only 5 localized ASKAP FRBs, which is consistent with the joint result of BBN and CMB. Nevertheless, the uncertainty in the result of Ref.~\citep{macquart2020census} is fairly large (at a $\sim 40\%$ precision), and thus it is necessary to investigate what extent the missing baryon problem can be solved to by using the SKA-era FRB observation. 

In this paper, we simulate the SKA-era FRB observation and investigate the important cosmological issues using the mock FRB data. Our focuses of scientific problems are on the issues of dark energy, the Hubble constant and baryons. 

This work is organized as follows. In Section \ref{sec2}, we introduce FRB and GW simulation methods. We present the constraint results and relevant discussions in Section \ref{sec3}. Conclusions are given in Section \ref{sec4}.

\section{Methods and data}\label{sec2}

\subsection{Uncertainties in dispersion measures of FRBs}\label{21}
When a FRB's emission travels through the plasma, the radio pulse will be dispersed, i.e. photons at higher frequency arrive earlier. The time delay ($\Delta t$) of frequencies $\nu_{1}$ and $\nu_{2}$ ($\nu_{1}<\nu_{2}$) can be quantified by DM  \citep{Petroff:2019tty},
\begin{equation}\label{eq23}
\begin{aligned} \Delta t \simeq 4.15 \rm{~s}\left[\left(\frac{\nu_{1}}{ \rm{GHz}}\right)^{-2}-\left(\frac{\nu_{2}}{\rm{GHz}}\right)^{-2}\right] \frac{\rm{DM}}{10^{3} ~\rm{pc} ~\rm{cm}^{-3}} \end{aligned}.
\end{equation}
The observed DM of a FRB at redshift $z$, representing the electron density integrated along the line-of-sight ($\rm{DM}=\int \emph{n}_{\rm{e}}\rm{~d} \emph{l}/(1+\emph{z})$), can be separated into the following components \citep{thornton2013pop,Deng:2013aga},
\begin{equation}\label{eq2}
\rm{DM}_{\rm{obs}}=\rm{DM}_{\rm{MW}}+\rm{DM}_{\rm{E}},
\end{equation}
where $\rm{DM}_{\rm{MW}}$ is the Galactic contribution from the interstellar medium (ISM) and halo of the Milky Way, and $\rm{DM}_{\rm{E}}$ is the extragalactic contribution. The component $\rm{DM}_{\rm{MW}}$ can be divided into the
ISM-contributed $\rm{DM}_{\rm{MW},\text {ISM}}$ and the halo-contributed $\rm{DM}_{\rm{MW},\text {halo}}$. Given the Galactic coordinate of a FRB, $\rm{DM}_{\rm{MW},\text {ISM}}$ can be obtained from typical electron density models such as NE2001 \citep{Cordes:2002wz} or YMW16 \citep{yao2017new}, while the latter can be estimated as $\rm{DM}_{\rm{MW}, \text{halo}}=50$--$80 ~{\rm {pc~cm^{-3}}}$ \citep{pro2019probing}.
The extragalactic DM of a FRB can be defined as
\begin{equation}\label{eq22}
\rm{DM}_{\rm{E}} \equiv\rm{DM}_{\rm{obs}}-\rm{DM}_{\rm{MW}}=\rm{DM}_{\rm{IGM}}+\frac{\rm{DM}_{\text {host}}+\rm{DM}_{\text {src}}}{1+\emph{z}},
\end{equation}
where $\rm{DM}_{\rm{IGM}}$ is the dominant component contributed by the IGM, and $\rm{DM}_{\rm{host}}$ and $\rm{DM}_{\rm{src}}$ represent the contributions from the host galaxy and the source's local environment, respectively. 
In particular, $\rm{DM}_{\rm{IGM}}$ is closely related to cosmology, and Macquart relation gives its averaged value, i.e.,
\begin{equation}\label{eq3}
\left\langle\rm{DM}_{\rm{IGM}}\right\rangle=\frac{3 c \Omega_{\rm b} H_{0}}{8 \pi G m_{p}} \int_{0}^{z} \frac{f_{e}(z')f_{\rm{IGM}}(z')(1+z')}{E(z')} d z',
\end{equation}
where $G$ is the Newton's constant, $m_{\rm{p}}$ is the proton mass, $f_{\rm{IGM}}\simeq0.83$ is the fraction of baryon mass in the IGM \citep{shull2012baryon} and $E(z)$ is the dimensionless Hubble parameter related to cosmological models (see Eq.~(\ref{eq:Ez}) for details). The electron fraction is $f_{e}(z)=Y_{\rm H}\chi_{\rm{e,H}}(z)+\frac{1}{2}Y_{\rm He}\chi_{\rm{e,He}}(z)$, where $Y_{\rm H}=3/4$ and $Y_{\rm He}=1/4$ are the mass fractions of hydrogen and helium, respectively, and $\chi_{\rm {e,H}}$ and $\chi_{\rm {e,He}}$ are the ionization fractions of hydrogen and helium, respectively. $f_{e}(z)$ varies with both $\chi_{\rm {e,H}}$ and $\chi_{\rm {e,He}}$ during the hydrogen and helium reionization. We set $f_{\rm{e}}= 0.875$ since both hydrogen and helium are fully ionized at $z < 3$ \citep{fan2006observational}. From Eq.~(\ref{eq22}), $\rm{DM}_{\rm{E}}$ is available for a localized FRB with $\rm{DM}_{\rm{obs}}$ and $\rm{DM}_{\rm{MW}}$ determined.
If $\rm{DM}_{\rm{host}}$ is treated properly, $\rm{DM}_{\rm{IGM}}$ can be measured and the corresponding uncertainty can be expressed as
\begin{equation}\label{eq6}
\sigma_{\rm{DM}_{\rm{IGM}}}=\left[\sigma_{\rm obs}^{2}+\sigma_{\rm MW}^{2}+\sigma_{\rm IGM}^{2}
+\left(\frac{\sigma_{\rm host,src}}{1+z}\right)^{2} \right]^{1/2}.
\end{equation}
The observational uncertainty $\sigma_{\rm {obs}}=0.5~{\rm {pc~cm^{-3}}}$ is derived from the current catalogs of known FRBs \citep{Petroff:2019tty,CHIMEFRB:2021srp}. With the ATNF pulsar catalog \citep{Manchester:2004bp},
the uncertainty of $\rm{DM}_{\rm{MW}}$ averages about $10~{\rm {pc~cm^{-3}}}$ for the pulses from high Galactic latitude $\left(|b|>10^{\circ}\right)$.
The uncertainty $\sigma_{\rm {IGM}}$ from the mean $\rm{DM}_{\rm{IGM}}$ of whole population, regarded as the sightline-to-sightline variance due to the effect of cosmic baryonic inhomogeneity in the IGM, has greater leverage on $\sigma_{\rm{DM}_{\rm{IGM}}}$ \citep{Walters:2017afr}.
McQuinn \citep{mcquinn2013locating} figured out three baryonic profile models to determine the deviation.
Here we consider the simplest one, namely, the top hat model of a putative $\sigma_{\rm IGM}$ scales with redshift,
\begin{equation}
\sigma_{\rm{IGM}}(z)=173.8~z^{0.4}~\rm{pc}~\rm{cm}^{-3},
\end{equation}
which is fitted in a power-law form \citep{Qiang:2020vta}. For an individual in FRB populations, $\sigma_{\rm{host,src}}$ is actually characterized by the distinctive properties, e.g., the type of the host galaxy, the site of FRB in the host, and the near-source plasma, leading to the difficulty for its estimation. Here we adopt $\sigma_{\rm{host, src}} = 30 ~{\rm {pc~cm^{-3}}}$ as the uncertainty of both $\rm{DM}_{\rm{host}}$ and $\rm{DM}_{\rm{src}}$. 

\subsection{Event rate of FRBs}\label{22}
Owing to the high all-sky FRB rate of $\sim 10^5$ per day \citep{Niu:2021xug}, a large number of FRBs are expected to be accumulated by the upcoming surveys. In order to predict the detectable sample size of localized FRB data in the SKA era, current samples are applied to perform statistical studies, e.g., on event rate distribution \citep{bera2016modelling,lawrence2017nonhomogeneous}, energy function \citep{Zhang:2019yiw,Hashimoto:2022llm,Zhang:2022rib,Li:2023zro} and volumetric event rate of the FRB population \citep{james2019limits}.

To count the cumulative event number of FRBs above a specific fluence threshold $F_{\nu}$, a power-law model $N\left(>F_{\nu}\right) \propto F_{\nu}^{\alpha}$ \citep{Macquart:2018jlq} is often used, where the power-law index $\alpha = -1.5$ is consistent with a non-evolving population in Euclidean space \citep{ved2016flu}.
The approach is instructive to estimate the all-sky event rate $N_{\rm sky}$ above a given fluence threshold $F_{\nu}$, which reads
\begin{equation} \label{fluence}
N_{\rm sky}(>F_{\nu})=N^{\prime}_{\rm sky}\left(\frac{F_{\nu}}{F^{\prime}_{\nu}}\right)^{\alpha} [\text {sky}^{-1} \text {d}^{-1}],
\end{equation}
where $N^{\prime}_{\rm sky}$ is the all-sky event rate at fluence threshold $F^{\prime}_{\nu}$. In other words, this method can extrapolate the target event rate $N_{\rm sky}$ from the known event rate $N^{\prime}_{\rm sky}$ estimated by existing FRB surveys. Notably in the simple power-law model, the measurements of $N_{\rm sky}$ will vary greatly with $\alpha$, which have been widely estimated in different surveys \citep{ved2016flu,James2018bys,agarwal2020ini,Zhang:2021kdu}. 

According to the suggestion of \citet{macquart2015fast}, we choose the mid-frequency array of the first phase of SKA (SKA1-MID) as the optimal array for FRB observation, which reaches a 10$\sigma$ fluence of $F_{\nu} =$ 14 mJy ms \citep{fender2015tra}. Here, we consider the rate measurements of ASKAP and Parkes telescopes \citep{staveley1996par}, which dominate at 1.4 GHz within the 0.9--1.67 GHz observing band of SKA1-MID. \citet{ANTARES2017hvn} obtain $N^{\prime}_{\rm sky} = 1.7_{-0.9}^{+1.5} \times 10^{3}$ sky$^{-1}$ d$^{-1}$ above the fluence of $F^{\prime}_{\nu} =$ 2 Jy ms from Parkes surveys. \citet{shannon2018dis} obtain $N^{\prime}_{\rm sky} = 37 \pm 8$ sky$^{-1}$ d$^{-1}$ above the fluence of $F^{\prime}_{\nu} =$ 26 Jy ms based on ASKAP FRB surveys. 

We simply adopt the Euclidean-expected value of $\alpha=-1.5$, although by analyzing the Parkes and ASKAP datasets some researches \citep{ved2016flu,James2018bys} report a slight disagreement with the Euclidean expectation. According to Eq.~(\ref{fluence}), we can calculate the all-sky rate observed by SKA1-MID based on the results of Parkes and ASKAP. Based on the Parkes result \citep{ANTARES2017hvn}, we have 
\begin{equation} 
\begin{aligned}
N_{\rm{sky}} &=(1.7_{-0.9}^{+1.5} \times 10^{3})\left(\frac{0.014~\rm{Jy}~\rm{ms}}{2~\rm{Jy} ~\rm{ms}}\right)^{-1.5} \text {sky}^{-1} \text {d}^{-1} \\
&= 2.90_{-1.54}^{+2.56} \times 10^{6} ~\text {sky}^{-1} \text {d}^{-1},
\end{aligned}
\end{equation}
and based on the ASKAP result \citep{shannon2018dis}, we have 
\begin{equation} 
\begin{aligned}
N_{\rm{sky}} &=(37 \pm 8)\left(\frac{0.014~\rm{Jy}~\rm{ms}}{26~\rm{Jy} ~\rm{ms}}\right)^{-1.5} \text {sky}^{-1} \text {d}^{-1} \\
&= (2.96 \pm 0.64) \times 10^{6} ~\text {sky}^{-1} \text {d}^{-1}.
\end{aligned}
\end{equation}
It is clearly seen that the two results are well consistent with each other. 

\begin{figure*}
\begin{minipage}[t]{0.49\textwidth}
\centering
\includegraphics[width=\linewidth,angle=0]{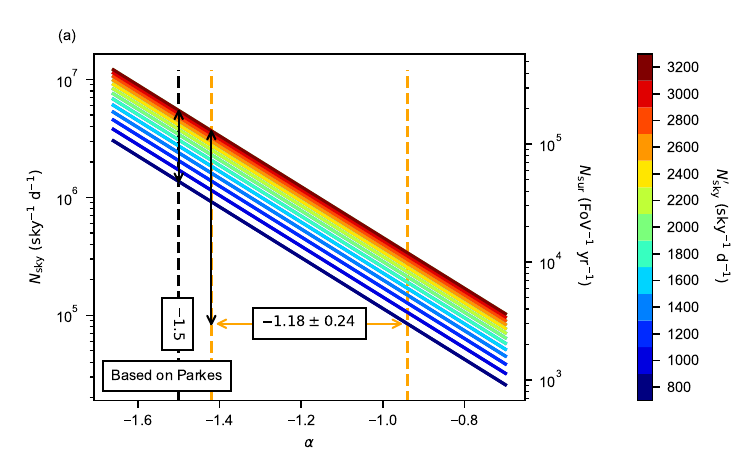}

\end{minipage} \hfill
\begin{minipage}[t]{0.49\textwidth}
\centering
\includegraphics[width=\linewidth,angle=0]{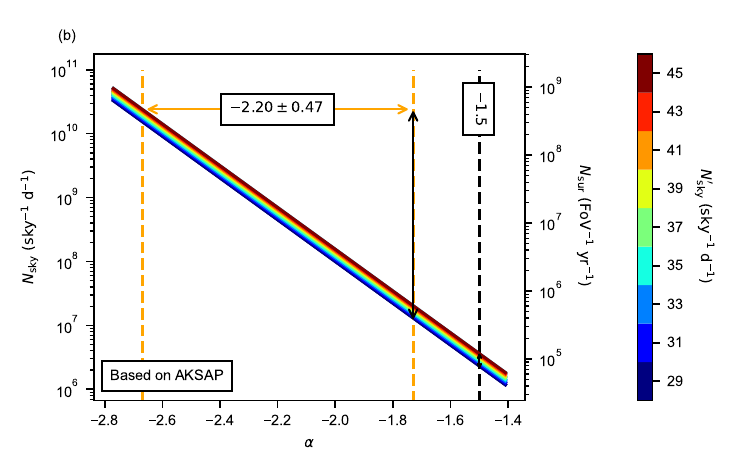}
\end{minipage}
\caption{The estimated all-sky event rates for SKA-MID in units of $\rm{sky}^{-1}~\rm{d}^{-1}$ (left vertical axis) and detection event rates in units of $\rm{FoV}^{-1}~\rm{yr}^{-1}$ (right vertical axis) as functions of $\alpha$. The coloured solid lines indicate the functions calculated at different all-sky event rates from Parkes (a) and ASKAP (b), respectively. The black vertical dashed lines denote the case of $\alpha = -1.5$, while the orange ones correspond to $\alpha = -1.18 \pm 0.24$ and $\alpha = -2.20 \pm 0.47$ from Ref.~\citep{James2018bys}. The black solid arrows cover the event-rate ranges for different $\alpha$ cases. The SKA-MID fluence threshold is $F_{\nu} = 0.014$ Jy ms.}\label{Fig.alpha} 
\end{figure*}

Then we can derive the apparent detection rate $N_{\rm{sur}}$ for a given telescope,
\begin{equation}\label{Nsurvey}
N_{\rm{sur}}=N_{\rm{sky}}[\rm{sky}^{-1}\rm{d}^{-1}]\Omega[\text{sky}~\text{FoV}^{-1}]t_{\rm{obs}} [\text {d}~\text {yr}^{-1}], 
\end{equation}
where $\Omega$ is the sky coverage fraction\footnote{Note that $\Omega$ is calculated as $\Omega = \Omega_{\text{sur}}/\Omega_{\text{tot}}$ in units of $\text{sky}~\text{FoV}^{-1}$ (or $\text{FoV}^{-1}/\text{sky}^{-1}$), where $\Omega_{\text{sur}}$ and $\Omega_{\text{tot}}$ stand for the solid angle of FoV and the whole sky, respectively.} of the SKA instantaneous observation, and $t_{\text{obs}}$ is the exposure time on source. SKA1-MID can be used to commensally search for FRBs along with pulsar searches. Equipped with phased array feeds (PAFs) for each dish to realize wide field of view (FoV), SKA1-MID has a $\sim 20$ $\rm{deg^2}$ effective instantaneous FoV \citep{CRAFT:2010dep}, which is used to calculate $\Omega$.\footnote{In general, the survey area of the SKA1-MID in Medium-Deep Band 2 can reach $\sim 5,000$ $\rm{deg^2}$ during the overall observation. Since FRB serves as transient phenomenon, here we take $\Omega_{\text{sur}} = 20~\rm{deg^2}$ to calculate the event rate of FRBs in the instantaneous FoV of the SKA1-MID equipped with PAFs.}
For $t_{\text{obs}}$, according to Ref.~\citep{Bhandari:2019med}, which introduced the strategy of a FRB search on ASKAP viewed as a precursor to the SKA, we take an average of $20\%$ of observing time per year spent observing FRBs, and take $t_{\text{obs}} = 20\% \times 365$ $\text {d}~\text {yr}^{-1}$. Thus, we can estimate the expected event number of SKA1-MID 
per year, i.e., $N_{\rm sur} \simeq 1.03_{-0.54}^{+0.91} \times 10^{5}~\rm{FoV}^{-1}~\rm{yr}^{-1}$ and $(1.05 \pm 0.23) \times 10^{5}~ \rm{FoV}^{-1}~\rm{yr}^{-1}$, which are the predictions based on the Parkes and ASKAP results, respectively. These results reach a consensus with previous assessments \citep{Fialkov:2017qoz,hashimoto2020fast}. We conclude that ${\cal O}(10^5)$--${\cal O}(10^6)$ FRBs can be detected by SKA1-MID in a 10-year observation. The estimated results are summarized in Table~\ref{tab:1}. We also discuss the influence of $\alpha$ on our results by employing the 1$\sigma$ estimates of $\alpha = -1.18 \pm 0.24$ and $\alpha = -2.20 \pm 0.47$ from Ref.~\citep{James2018bys}. Based on Parkes and ASKAP event rates, we can plot $N_{\rm sky}$ (or $N_{\rm sur}$) as functions of $\alpha$, as shown in the left and right panels of Figure~\ref{Fig.alpha}, respectively. We find that $\alpha$ is inversely proportional to $N_{\rm sky}$ for every fixed $N^{\prime}_{\rm sky}$. Therefore, the variation of $\alpha$ can increase the ASKAP-based result one to three orders of magnitude, while reducing the Parkes-based result by up to one order of magnitude. 

\begin{table*}[!htbp]
\centering
\caption{Summary of all-sky and detection event rates for SKA-MID at fluence threshold $F_{\nu} = 0.014$ Jy ms. }
\label{tab:1}
\resizebox{\textwidth}{!}{%
\vspace{0.5cm}
\setlength{\tabcolsep}{3mm}
\renewcommand{\arraystretch}{2.5
}

\begin{tabular}{ccccc}
\hline \hline
\multicolumn{2}{c}{FRB survey} & Power-law index & All-sky event rate   ($\rm{sky}^{-1}~\rm{d}^{-1}$) & Detection event rate   ($\rm{FoV}^{-1}~\rm{yr}^{-1}$) \\ \hline
\multirow{4}{*}{SKA-MID} &
  \multirow{2}{*}{Based on Parkes} &
  $-1.5$ &
  $2.90_{-1.54}^{+2.56} \times 10^{6}$ &
  $1.03_{-0.54}^{+0.91} \times 10^{5}$ \\
 &
   &
  $-1.18 \pm 0.24$ \citep{James2018bys} &
  $2.90_{-2.82}^{+0.77} \times 10^{6}$ &
  $1.03_{-1.00}^{+0.27} \times 10^{5}$ \\ \cline{2-5} 
 &
  \multirow{2}{*}{Based on ASKAP} &
  $-1.5$ &
  $(2.96 \pm 0.64) \times 10^{6}$ &
  $(1.05 \pm 0.23) \times 10^{5}$ \\
 &
   &
  $-2.20 \pm 0.47$  \citep{James2018bys} &
  $\mathcal{O}(10^7)$--$\mathcal{O}(10^{10})$ &
  $\mathcal{O}(10^5)$--$\mathcal{O}(10^{8})$ \\ \hline \hline
\end{tabular}%
}

\end{table*}

The event localization of FRBs is paramount for FRB cosmology. To identify the host galaxies and hence redshifts, it is necessary to localize FRBs within $\sim 0.1''-0.5''$ \citep{macquart2015fast}. Within an accuracy of a few sub-arcseconds, ASKAP has localized at least 16 FRBs in dozens of its detected samples \citep{James:2022dcx}. The SKA is capable of localizing FRBs to the sub-arcsecond level, and thus it is reasonable to assume that all the FRBs detected by the SKA can be simultaneously localized (and their redshifts can also be measured).

\subsection{Uncertainties in luminosity distances of GW standard sirens}
\label{GW}
{We simulate the GW standard sirens based on the Einstein Telescope (ET) \citep{Punturo:2010zz}, considering only the binary neutron star (BNS) merger events. The redshift distribution of BNSs is expressed as \citep{zhao2011det,Cai:2016sby}
\begin{align}
P(z)\propto \displaystyle{\frac{4\pi d_{\rm C}^2(z)R(z)}{H(z)(1+z)}},
\end{align}
where $d_{\rm C}$ is the comoving distance and $R(z)$ is the time evolution of the burst rate with the form \citep{Schneider:2000sg,Cutler:2009qv,Cai:2016sby}
\begin{eqnarray}
R(z) =
\begin{cases}
1+2z, &z\leq 1, \\
\displaystyle{\frac{3}{4}}(5-z), &  1 < z < 5,\\
0,  &  z \geq 5.
\end{cases}
\end{eqnarray}
The amplitude of the GW signal in Fourier space $\mathcal{A}$ and the luminosity distance $d_{\rm{L}}$ are approximately  related by $\mathcal{A} \propto 1/d_{\rm L}$. The luminosity distance in a spatially-flat universe is expressed as 
\begin{equation}\label{dl}
d_{\rm{L}}(z)=(1+z) d_{\rm{C}}(z)=(1+z) \int_{0}^{z} \frac{d z^{\prime}}{H(z^{\prime})},
\end{equation}
where $H(z)$ is the Hubble parameter.}

{The total errors in the measurement of $d_{\rm L}$ include the instrumental, weak lensing, and peculiar velocity errors, i.e.,
\begin{align}\label{sigdl}
\sigma_{d_{\rm L}}=\sqrt{(\sigma_{d_{\rm L}}^{\rm inst})^2 + (\sigma_{d_{\rm L}}^{\rm lens})^2 + (\sigma_{d_{\rm L}}^{\rm pv})^2}.
\end{align}
The instrumental error is simulated using the method described in Ref.~\citep{Nishizawa:2010xx}, the weak lensing and peculiar velocity errors are obtained from Refs.~\citep{Hirata:2010ba} and \citep{Gordon:2007zw}, respectively. Assuming a 10-year observation span, the ET is expected to detect 1000 GW events from BNS mergers in the redshift range of $z\lesssim5$ \citep{Jin:2020hmc,Zhang:2019loq}. In this work, we simulate this catalog. Other studies on the GW standard sirens from the coalescences of (super)massive black hole binaries based on various space-based GW observatories and pulsar timing arrays can be found in Refs.~\citep{Wang:2019tto,Zhao:2019gyk,Wang:2021srv,TianQin:2020hid,LISACosmologyWorkingGroup:2022jok,Wang:2022oou}.}

\subsection{Fiducial cosmological models}
\label{Cd} 
The dark-energy EoS parameter is defined as $w(z)=p_{\rm de}(z)/\rho_{\rm de}(z)$, with $p_{\rm de}(z)$ and $\rho_{\rm de}(z)$ pressure and density of dark energy, respectively. In a spatially-flat universe, the dimensionless Hubble parameter is given by the Friedmann equation,
\begin{align}\label{eq:Ez}
E^2(z)=\frac{H^2(z)}{H_0^2}=\,&(1 - {\Omega _{\rm m}})\exp \left[3\int_0^z {\frac{{1 + w(z')}}{{1 + z'}}} d z'\right] \nonumber\\
&+{\Omega _{\rm m}}{(1 + z)^3},
\end{align}
where ${\Omega_{\rm m}}$ is the present-day matter density parameter.

In this work, we consider three fiducial cosmological models to generate simulated data of FRBs (the central values of these data). We only consider the simplest and most general cases in the studies of dark energy, namely, the $\Lambda$CDM model [$w (z) = -1$] in which the cosmological constant $\Lambda$ (equivalent to the vacuum energy density) serves as dark energy, the $w$CDM model [$w(z)$ = constant] in which dark energy has a constant EoS, and the $w_{0}w_{a}$CDM model [$w(z)=w_{0}+w_{a}z/(1+z)$] \citep{chevallier2001accelerating,linder2003exploring} in which dark energy has an evolving EoS characterized by two parameters $w_0$ and $w_a$. We use the simulated FRB data and other observational data to constrain the three dark energy models.


\subsection{Other observational data}

In this work, we also use other observational data as complementary. In addition to the CMB data, we also use the mainstream late-universe observations, BAO and SNe, to combine with CMB. 

For the CMB observation, we employ the ``Planck distance priors" derived from Planck 2018 observation \citep{chen2019distance}.  
We did not use the full power spectra data of CMB, because the actual observational data are only used as complementary in this work and utilization of CMB distance prior data is convenient and resource saving. 

For the BAO data, we use five data points from three observations, including the measurements from 6dF Galaxy Survey (6dFGS) at $z_{\rm eff} = 0.106$ \citep{beutler20116df}, Main Galaxy Sample of Data Release 7 of Sloan Digital Sky Survey (SDSS-MGS) at $z_{\rm eff} = 0.15$ \citep{ross2015clustering}, and Data Release 12 of Baryon Oscillation Spectroscopic Survey (BOSS-DR12) at $z_{\rm eff} = 0.38$, 0.51 and 0.61 \citep{alam2017clustering}. For the SNe data, we use the latest Pantheon data set compiled by  \citet{scolnic2018complete}. 

We use the data combination CMB+BAO+SNe (abbreviated as ``CBS") to constrain the fiducial cosmological models, and the obtained best-fit cosmological parameters that are used to generate the central values of the simulated FRB data.

Moreover, the local-universe measurement result of $H_{0}= 73.04 \pm 1.04 \rm{~km} \rm{~s}^{-1} \rm{Mpc}^{-1}$ from cosmic distance ladder given by the SH0ES team \citep{Riess:2021jrx} is employed as a Gaussian prior when we discuss the missing baryon problem.

\subsection{Method of cosmological parameter estimation}\label{Se}

\begin{figure*}[htb]
\includegraphics[width=0.9\linewidth,angle=0]{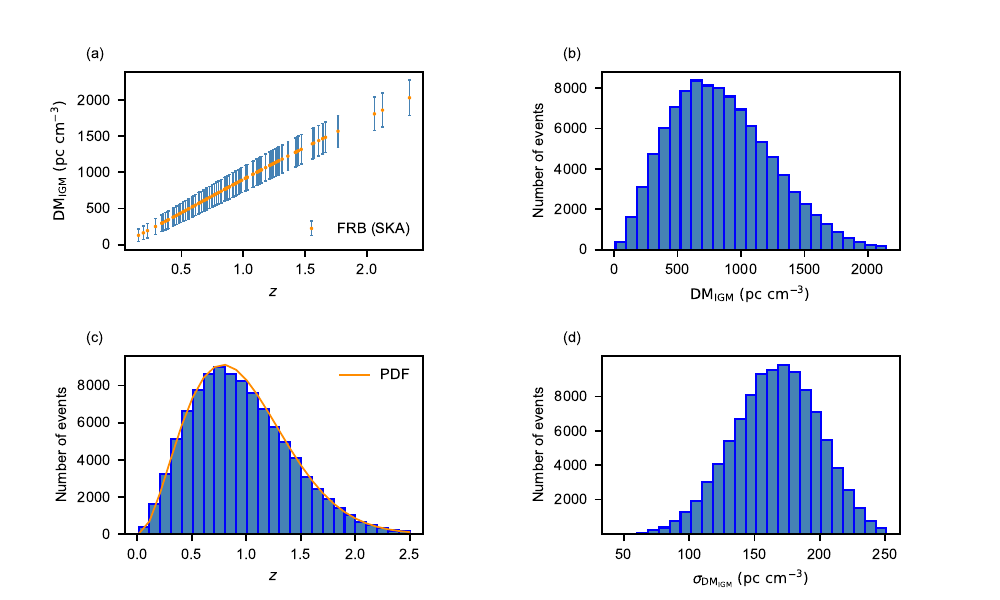}
\caption{The simulated FRB data (FRB1 with $N_{\rm{FRB}} = 10^{5}$) for a 10-year observation of the SKA based on the fiducial $\Lambda$CDM cosmology. (a) The data points of ${\rm DM}_{\rm IGM}(z)$ with $1\sigma$ errors. Here, one point represents 1000 events. The orange points are the central values and the blue bars are the $1\sigma$ uncertainties. This plot directly shows the Macquart relation (see Eq.~(\ref{eq3})). (b) The histogram of FRB event number versus ${\rm DM}_{\rm IGM}$. (c) The histogram of FRB event number versus $z$. Here the probability distribution function (PDF) of redshift is also plotted as the orange line to show the constant-density redshift distribution function (see Eq.~(\ref{Nconst})) used in this work. (d) The histogram of FRB event number versus $\sigma_{\rm DM_{IGM}}$.
}\label{Fig1}
\end{figure*}

Currently, the real redshift distribution of FRBs is still unknown. Several studies investigated this distribution using current different FRB databases \citep{Zhang:2021kdu,James:2021oep,Zhang:2020ass,James:2021jbo}, but no consistent result was obtained due to the limited number of detected FRBs. \citet{Qiang:2021bwb} highlighted the significant effect of the assumed redshift distributions on cosmological parameter estimation, leading to different scenarios for the prospects of FRB cosmology. In this work, we assume a constant comoving number density distribution with a Gaussian cut-off $z_{\rm cut}=1$, which makes FRBs have a moderate constraining capability as suggested in Ref.~\citep{Qiang:2021bwb}. The constant-density distribution function $N_{\text {const}}(z)$ is expressed as
\begin{equation}
\label{Nconst}
N_{\text {const}}(z)=\mathcal{N}_{\text {const}} \frac{d_{\rm{C}}^{2}(z)}{H(z)(1+z)} e^{-d_{\rm{L}}^{2}(z) /\left[2 d_{\rm{L}}^{2}\left(z_{\rm{cut}}\right)\right]},
\end{equation}
where $\mathcal{N}_{\text{const}}$ is a normalization factor.

We generate the mock data of FRBs in the redshift range of $0<z<2.5$ tracking the aforementioned distribution. The central values of DMs in the simulated data of FRBs are given by the fiducial models, i.e. the $\Lambda$CDM, $w$CDM and $w_{0}w_{a}$CDM models, that are constrained by the CBS data. The uncertainties in DMs are considered according to the prescription given in Section~\ref{21}. We carry out the Markov Chain Monte Carlo (MCMC) analysis \citep{lewis2002cosmological} to derive the posterior probability distributions of cosmological parameters. 

So we use the $\chi^2$ statistic as a measure of likelihood for the parameter values. The $\chi^2$ function of FRB is defined as
\begin{equation}
\chi_{\rm{FRB}}^{2}(\boldsymbol{\theta})=\sum_{i=1}^{N_{\rm FRB}}\left(\frac{{\rm{DM}_{\rm{IGM}}^{\rm{th}}}(z_{i}; {\boldsymbol{\theta}})-{\rm{DM}_{\rm{IGM}}^{\rm{obs}}}(z_{i})}{\sigma_{\rm{DM}_{\rm{IGM}}}(z_{i})}\right)^{2},
\end{equation}
where $\boldsymbol{\theta}$ represents the set of cosmological parameters, ${\rm DM}_{\rm{IGM}}^{\rm{th}}(z_i; \boldsymbol{\theta})$ is the theoretical value of $\rm{DM}_{\rm{IGM}}$ at the redshift $z_i$ calculated by Eq.~(\ref{eq3}), 
$\rm{DM}_{\rm{IGM}}^{\text {obs}}$($z_{i}$) is the observable value, and $\sigma_{\rm{DM}_{\rm{IGM}}}$($z_{i}$) represents the uncertainty of $\rm{DM}_{\rm{IGM}}$ (see Eq.~(\ref{eq6})).
{The $\chi^2$ function of GW is defined
as
\begin{equation}
\chi_{\rm{GW}}^{2}(\boldsymbol{\theta})=\sum_{i=1}^{N_{\rm GW}}\left(\frac{d_{\rm{L}}^{\rm{th}}(z_{i}; {\boldsymbol{\theta}})-d_{\rm{L}}^{\rm{obs}}(z_{i})}{\sigma_{d_{\rm L}}(z_{i})}\right)^{2},
\end{equation}
where $d_{\rm{L}}^{\rm{th}}(z_{i}; {\boldsymbol{\theta}})$ is the theoretical value of $d_{\rm{L}}$ at the redshift $z_i$ calculated by Eq.~(\ref{dl}), 
$d_{\rm{L}}^{\rm{obs}}$($z_{i}$) is the observable value, and $\sigma_{d_{\rm L}}$($z_{i}$) represents the uncertainty of $d_{\rm{L}}^{\rm{th}}$ (see Eq.~(\ref{sigdl}))}

In our work, the Python package {\tt emcee} \citep{foreman2013emcee} is employed for the MCMC analysis, and {\tt GetDist} \cite{getdistweb} is used for plotting the posterior distributions of the cosmological parameters.

We consider a 10-year observation of the SKA for the detections of FRBs. According to our estimation of the FRB detection event rate by the SKA in Section~\ref{22}, the 10-year observation of the SKA would detect ${\cal O}(10^5-10^6)$ FRBs. Thus, in this work, we consider two cases: the normal expected scenario with $N_{\rm{FRB}} = 10^{5}$ (denoted as FRB1) and the optimistic expected scenario $N_{\rm{FRB}} = 10^{6}$ (denoted as FRB2). The simulated FRB1 data as an example is shown in Figure \ref{Fig1}.

\section{Results and discussion}\label{sec3}

\begin{table*}[!htbp]
\caption{The absolute errors (1$\sigma$) and the relative errors on the cosmological parameters in the $\Lambda$CDM, $w$CDM and $w_{0}w_{a}$CDM models using the CMB, CBS, FRB1, CMB+FRB1, FRB2 and CMB+FRB2 data. Here CBS stands for CMB+BAO+SNe, and FRB1 and FRB2 denote the simulated FRB data by a 10-year observation of the SKA in the normal expected scenario ($N_{\rm{FRB}} = 10^{5}$) and the optimistic expected scenario ($N_{\rm{FRB}} = 10^{6}$), respectively. $H_0$ is in units of km s$^{-1}$ Mpc$^{-1}$.}
\label{tab:CMB}
\setlength{\tabcolsep}{3mm}
\renewcommand{\arraystretch}{1.5}
\begin{center}{\centerline{
\begin{tabular}{ccm{1cm}<{\centering}m{1cm}<{\centering}m{1cm}<{\centering}m{1.5cm}<{\centering}m{1.5cm}<{\centering}m{1cm}<{\centering}m{1.5cm}<{\centering}m{1.5cm}<{\centering}
}
\hline \hline
Model & Error 
& CMB & CBS 
& FRB1 & CMB+FRB1 & CBS+FRB1 
& FRB2 & CMB+FRB2 & CBS+FRB2\\
\hline
\multirow{4}{*}{$\Lambda$CDM}
& $\sigma(\Omega_{\rm m})$  
& 0.0086 & 0.0059 
& 0.0036 & 0.0033 & 0.0030 
& 0.0012 & 0.0011 & 0.0011\\
& $\sigma(H_{0})$   
& 0.620 & 0.430 
& ---   & 0.240 & 0.210 
& ---   & 0.092 & 0.088\\
& $\varepsilon(\Omega_{\rm m})$ 
& $2.7\%$ & $1.9\%$ 
& $1.2\%$ & $1.1\%$ & $1.0\%$
& $0.4\%$ & $0.4\%$ & $0.4\%$ \\
& $\varepsilon(H_{0})$   
& $0.9\%$    & $0.6\%$ 
& ---        & $0.4\%$  & $0.3\%$
& ---        & $0.1\%$  & $0.1\%$ \\
\hline
\multirow{6}{*}{$w$CDM}
& $\sigma(\Omega_{\rm m})$  
& 0.0430  & 0.0078  
& 0.0043  & 0.0033 & 0.0030 
& 0.0013  & 0.0012 & 0.0012 \\
& $\sigma(H_{0})$   
& 4.40    & 0.83 
& ---     & 0.35   & 0.32 
& ---     & 0.14   & 0.14\\
& $\sigma(w)$   
& 0.150   & 0.035  
& 0.051   & 0.022  & 0.020 
& 0.016   & 0.013  & 0.012\\
& $\varepsilon(\Omega_{\rm m})$  
& $13.4\%$   & $2.5\%$ 
& $1.4\%$    & $1.1\%$   & $1.0\%$ 
& $0.4\%$    & $0.4\%$   & $0.4\%$\\
& $\varepsilon(H_{0})$   
& $6.5\%$    & $1.2\%$  
& ---        & $0.5\%$ & $0.5\%$ 
& ---        & $0.2\%$ & $0.2\%$\\
& $\varepsilon(w)$  
& $15.2\%$ & $3.5\%$  
& $5.0\%$  & $2.2\%$ & $2.0\%$ 
& $1.6\%$  & $1.3\%$ & $1.2\%$\\
\hline
\multirow{7}{*}{$w_{0}w_{a}$CDM}
& $\sigma(\Omega_{\rm m})$
& 0.0500  & 0.0079
& 0.0330  & 0.0068 & 0.0053
& 0.0120  & 0.0042 & 0.0038 \\
& $\sigma(H_{0})$   
& 5.50    & 0.84
& ---     & 0.70   & 0.49
& ---     & 0.27   & 0.25 \\ 
& $\sigma(w_{0})$   
& 0.440   & 0.083 
& 0.140   & 0.080  & 0.055
& 0.053   & 0.030  & 0.028\\
& $\sigma(w_{a})$  
& 1.43   & 0.32
& 0.82   & 0.21   & 0.16
& 0.31   & 0.12   & 0.11\\
& $\varepsilon(\Omega_{\rm m})$ 
& $15.9\%$ & $2.5\%$
& $10.7\%$ & $2.2\%$ & $1.7\%$
& $3.9\%$  & $1.4\%$ & $1.2\%$\\
& $\varepsilon(H_{0})$  
& $8.1\%$  & $1.1\%$
& ---    & $1.0\%$   & $0.7\%$
& ---    & $0.4\%$   & $0.4\%$\\
& $\varepsilon(w_{0})$ 
& $51.8\%$  & $8.3\%$ 
& $14.1\%$  & $7.6\%$  & $5.5\%$
& $5.3\%$   & $3.0\%$  & $2.8\%$\\
\hline \hline
\end{tabular}}}
\end{center}
\end{table*}

\begin{table*}[!htbp]
\caption{The absolute errors (1$\sigma$) and the relative errors on the cosmological parameters in the $\Lambda$CDM, $w$CDM and $w_{0}w_{a}$CDM models using the GW, FRB1, GW+FRB1, FRB2 and GW+FRB2 data. Here GW denotes the simulated GW standard siren data by a 10-year observation of ET.}
\label{tab:GW}
\setlength{\tabcolsep}{3mm}
\renewcommand{\arraystretch}{1.5}
\begin{center}{\centerline{
\begin{tabular}{ccm{2cm}<{\centering}m{2cm}<{\centering}m{3cm}<{\centering}m{2cm}<{\centering}m{3cm}<{\centering}}
\hline \hline
Model & Error 
& GW 
& FRB1 & GW+FRB1 
& FRB2 & GW+FRB2 \\
\hline
\multirow{4}{*}{$\Lambda$CDM}
& $\sigma(\Omega_{\rm m})$  
& 0.0130  
& 0.0036 & 0.0035
& 0.0012 & 0.0011 \\
& $\sigma(H_{0})$   
& 0.52 
& --- & 0.24
& --- & 0.21 \\
& $\varepsilon(\Omega_{\rm m})$  
& $4.1\%$ 
& $1.2\%$ & $1.1\%$
& $0.4\%$ & $0.4\%$ \\
& $\varepsilon(H_{0})$  
& $0.8\%$ 
& ---     & $0.4\%$
& ---     & $0.3\%$ \\
\hline
\multirow{6}{*}{$w$CDM}
& $\sigma(\Omega_{\rm m})$  
& 0.0190  
& 0.0043 & 0.0042 
& 0.0013 & 0.0013 \\
& $\sigma(H_{0})$   
& 1.20 
& --- & 0.51 
& --- & 0.26 \\
& $\sigma(w)$   
& 0.140  
& 0.051 & 0.046 
& 0.016 & 0.016 \\
& $\varepsilon(\Omega_{\rm m})$  
& $6.0\%$ 
& $1.4\%$ & $1.4\%$
& $0.4\%$ & $0.4\%$ \\
& $\varepsilon(H_{0})$  
& $1.8\%$ 
& ---     & $0.8\%$
& ---     & $0.4\%$ \\
& $\varepsilon(w)$  
& $13.7\%$ 
& $5.0\%$ & $4.6\%$
& $1.6\%$ & $1.6\%$ \\
\hline
\multirow{7}{*}{$w_{0}w_{a}$CDM}
& $\sigma(\Omega_{\rm m})$
& 0.046
& 0.033 & 0.030 
& 0.012 & 0.011  \\
& $\sigma(H_{0})$   
& 1.60
& ---   & 1.03 
& ---   & 0.48 \\ 
& $\sigma(w_{0})$   
& 0.220 
& 0.140 & 0.121 
& 0.053 & 0.051 \\
& $\sigma(w_{a})$  
& 1.29
& 0.82  & 0.70
& 0.31  & 0.30  \\
& $\varepsilon(\Omega_{\rm m})$  
& $14.6\%$ 
& $10.7\%$ & $9.6\%$
& $3.9\%$  & $3.5\%$ \\
& $\varepsilon(H_{0})$  
& $2.4\%$ 
& ---     & $1.5\%$
& ---     & $0.7\%$ \\
& $\varepsilon(w_0)$  
& $23.2\%$ 
& $14.1\%$ & $12.3\%$
& $5.3\%$  & $5.1\%$ \\
\hline \hline
\end{tabular}}}
\end{center}
\end{table*}

We use the simulated FRB observation of the SKA to constrain the cosmological models. In addition, to make comparison and combination, we also use the actual CMB and CBS data. For the priors on $\Omega_{\rm m}$, $H_0$ and $\Omega_{\rm b}h^{2}$, we use uniform distributions within ranges of $[0, 0.7]$, $[50, 90]$ $\rm km\ s^{-1}\ Mpc^{-1}$ and $[0, 0.05]$, respectively. We also adopt priors on $w$, $w_0$ and $w_a$ within ranges of $[-2, 1]$, $[-2, 2]$ and $[-3, 3]$ for dynamical dark energy models. The constraint results are summarized in Table~\ref{tab:CMB}. Here, we use $\sigma(\xi)$ and $\varepsilon(\xi)=\sigma(\xi)/\xi$ to represent the absolute and relative errors of the parameter $\xi$, respectively. In the following, we report the constraints on dark-energy parameter, the Hubble constant and baryon density in order.

\subsection{Dark energy}

\begin{figure*}
\begin{minipage}[t]{0.49\textwidth}
\centering
\includegraphics[width=0.8\linewidth,angle=0]{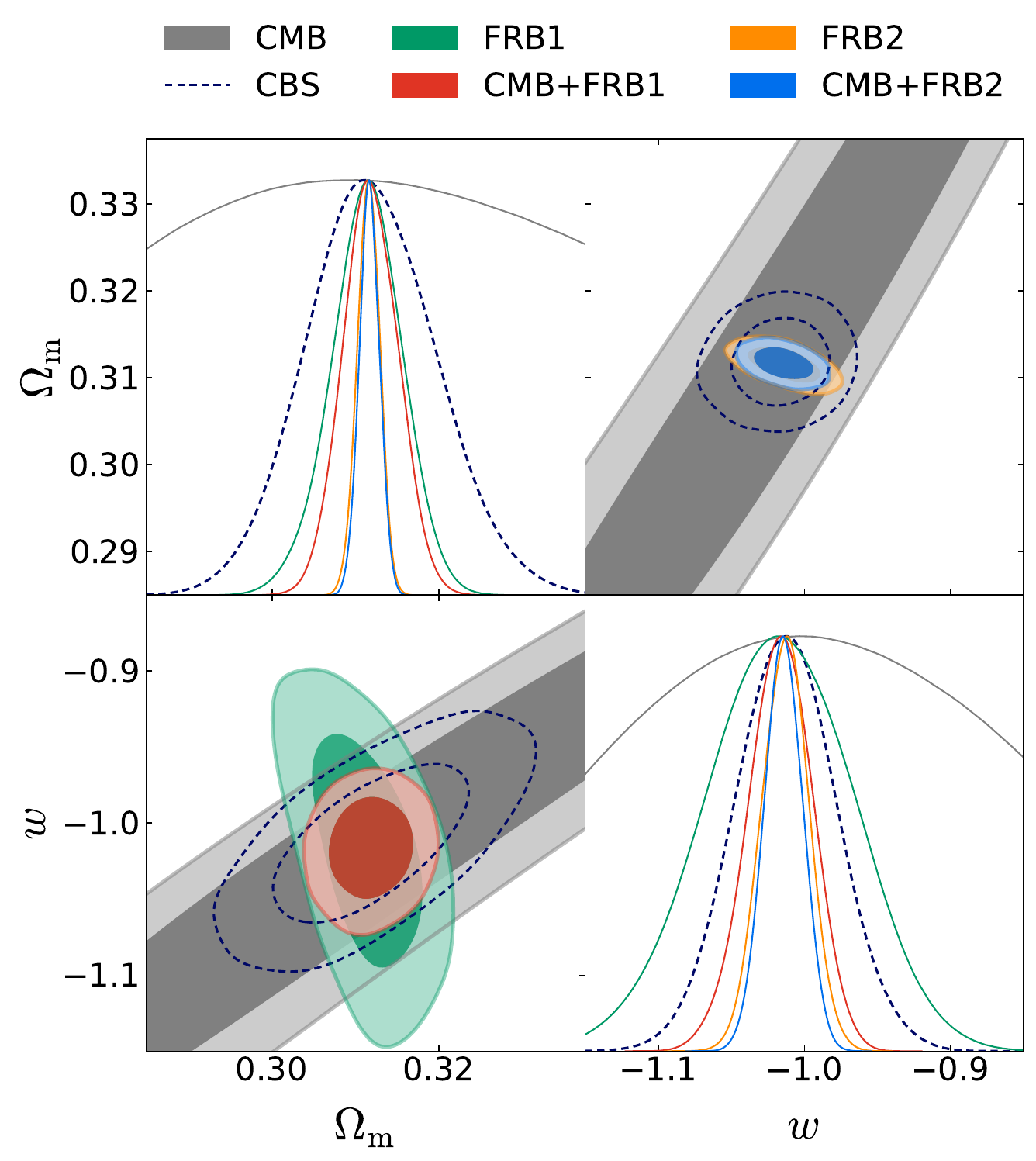}
\caption{Constraints ($68.3\%$ and $95.4\%$ confidence level) on $\Omega_{\rm m}$ and $w$ for the $w$CDM model by using the CMB, FRB1, CMB+FRB1, FRB2 and CMB+FRB2 data. For comparison, the dashed contour of CBS is also added.}\label{Fig.wcdm}                            
\end{minipage} \hfill
\begin{minipage}[t]{0.49\textwidth}
\centering
\includegraphics[width=0.8\linewidth,angle=0]{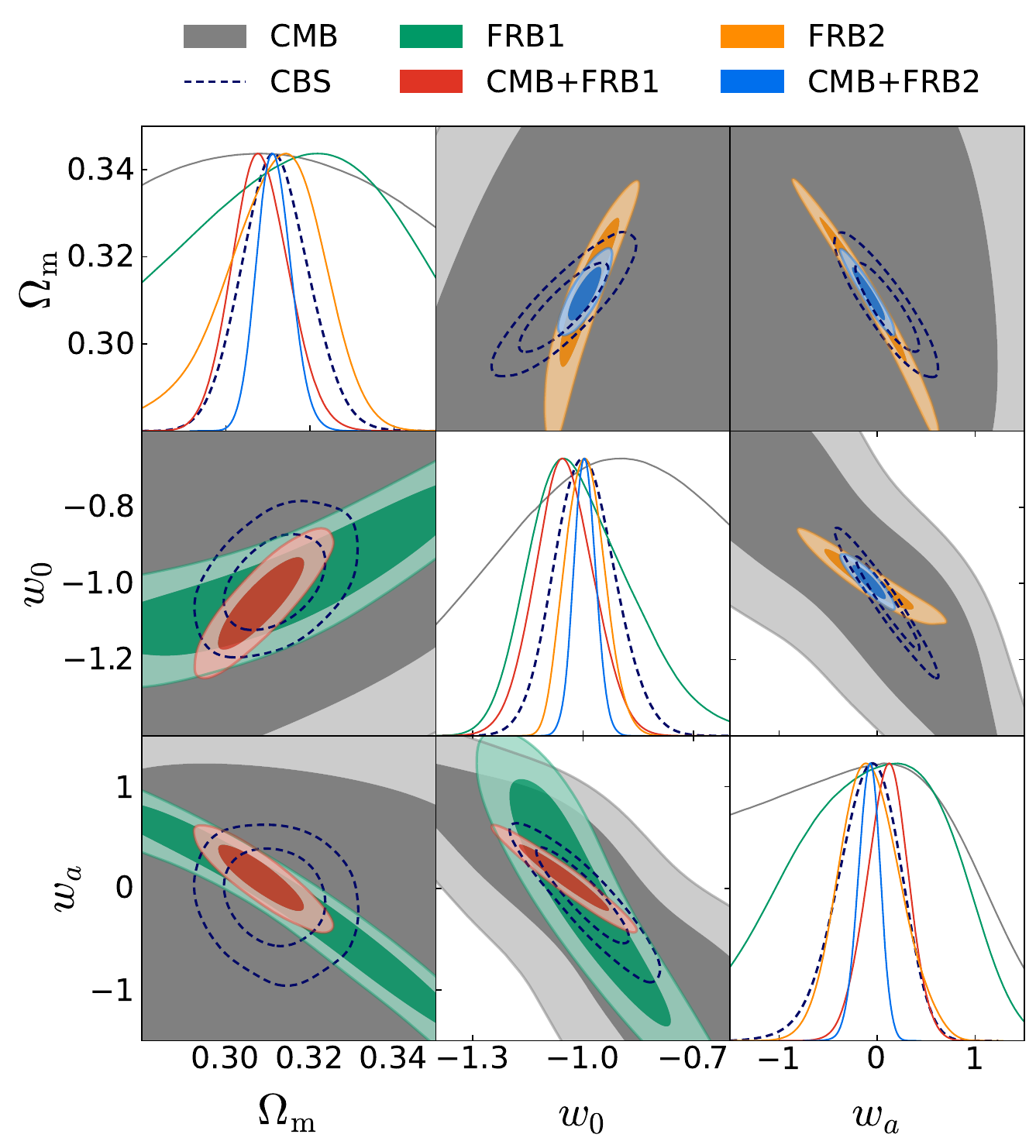}
\caption{Same as Figure~\ref{Fig.wcdm} but for the constraints on $\Omega_{\rm m}$, $w_0$ and $w_a$ for the $w_{0}w_{a}$CDM model.}\label{Fig.w0wa}
\end{minipage}
\end{figure*}

First, we discuss the constraints on the EoS of dark energy by using the FRB data. In Figs.~\ref{Fig.wcdm} and \ref{Fig.w0wa}, we show our constraint results on dark-energy EoS parameters for the $w$CDM and $w_{0}w_{a}$CDM models, respectively.

For the $w$CDM model, the CMB data alone cannot provide a precise constraint on the EoS of dark energy. For example, the CMB data can only give a $15.2\%$ constraint on $w$. By combining the CMB data with the current late-universe observations, BAO+SNe, we can see that the CBS data can lead to a $3.5\%$ constraint. 

However, we find that the FRB data alone can provide fairly tight constraints on $w$. Concretely, FRB1 and FRB2 give constraints of the $5.0\%$ and $1.6\%$ precision on $w$, respectively. Therefore, using about $10^5$ FRB data can provide a slightly weaker constraint than the CBS data, and using about $10^6$ FRB data can give a much better measurement on $w$ than the CBS data.

On combining the CMB data with the FRB1 data, similar to the findings in Ref.~\citep{zhao2020cosmological}, we find that this combination can effectively improve the constraints on dark-energy EoS parameters, compared to those solely using either the CMB or FRB data alone. For example, the combination of CMB+FRB1 provides a constraint of $\sigma(w)=0.022$, which is a significant improvement of about $85\%$ and $57\%$ compared to using the CMB and FRB1 data alone, respectively. The lower-left panel of Figure~\ref{Fig.wcdm} shows the constraint contours of CMB, FRB1, CMB+FRB1 in the $\Omega_{\rm m}$--$w$ plane. The orientations of the contours constrained by CMB and by FRB1 are rather different, indicating that FRB1 can well break the parameter degeneracy induced by CMB. In order to study the extent of this capability, we add the dashed contour of CBS in the figure for comparison with the contour of CMB+FRB1. It is evident that the CMB+FRB1 data can give better constraints than the CBS data, with a quantitative increase in precision of about $37\%$. This  suggests that using about $10^5$ FRB data can effectively break the parameter degeneracies induced by the CMB data, even better than using the data BAO+SNe.

When the number of FRB data is increased from $10^5$ (FRB1) to $10^6$ (FRB2), the FRB2 data alone can give a very tight constraint of $\sigma(w)=0.016$, while the CMB+FRB2 data can only give a slightly improved constraint of $\sigma(w)=0.013$ when compared to FRB2 alone. The upper-right panel of Figure~\ref{Fig.wcdm} shows the two-dimensional posterior contours by using CMB, FRB2, CMB+FRB2 and also CBS in the $\Omega_{\rm m}$--$w$ plane. The contour from FRB2 is significantly smaller than that of CMB, indicating that the constraint capability from CMB+FRB2 mainly arises from FRB2. By comparing with CBS, we can also see that FRB2 not only gives much better constraints than CMB but even better than CBS. This result once again highlights that, even without the help of CMB, using about $10^6$ FRB data can provide a remarkable constraint on $w$ independently, even tighter than the CBS data.

For the $w_{0}w_{a}$CDM model, the CMB data can only give a result of $\sigma(w_{0})=0.440$, which can be greatly improved to $\sigma(w_{0})=0.083$ by the CBS data. For $w_a$, it is poorly constrained by the CMB data but can be constrained by the CBS data to $\sigma(w_{a})=0.320$.  

We find that using about $10^5$ and $10^6$ FRB data can also provide tight constraints on both $w_{0}$ and $w_{a}$. The FRB1 data can give the absolute errors $\sigma(w_{0}) = 0.14$ and $\sigma(w_{a}) = 0.82$. Furthermore, the FRB2 data can give $\sigma(w_{0}) = 0.053$ and $\sigma(w_{a}) = 0.31$. Hence, we can see that using about $10^6$ FRB data can provide constraints on $w_{0}$ and $w_{a}$ that are comparable to those from the CBS data.

In our joint data analysis of the CMB and FRB data in the $w_{0}w_{a}$CDM model, the inclusion of the FRB data also can effectively break the parameter degeneracy formed by CMB. For example, CMB+FRB1 and CMB+FRB2 can significantly improve constraints on $w_0$ by $82\%$ and $93\%$, respectively, compared to using CMB alone. The bottom-middle panel of Figure~\ref{Fig.w0wa} shows the constraint contours of CMB, FRB1, CMB+FRB1 in the $w_0$--$w_a$ plane, and the center-right panel shows those of CMB, FRB2 and CMB+FRB2. For comparison, the dashed contour of CBS is also added in the figure. We find that, the constraint on $w_a$ from CMB+FRB1 is about $34\%$ better than CBS, and CMB+FRB2 is definitely better than CBS for the constraints both on $w_0$ and $w_a$, because FRB2 alone has the constraining capability comparable to CBS as previously reported. Concretely, CMB+FRB2 gives $\sigma(w_0) = 0.030$ and $\sigma(w_a) = 0.12$, which are both about $60\%$ better than CBS.

On combining the CBS data with the FRB data both in the $w$CDM and $w_{0}w_{a}$CDM models, we find that, in most cases (with the exception of CBS+FRB1 in the $w_{0}w_{a}$CDM model), the errors of constraints are similar with those from combining the CMB data. This suggests that, if the combination FRB+BAO+SNe is considered as a late-universe cosmological probe, the constraint capability is mainly from the FRB data of the SKA.

\begin{figure*}[!htbp]
\includegraphics[width=0.85\linewidth,angle=0]{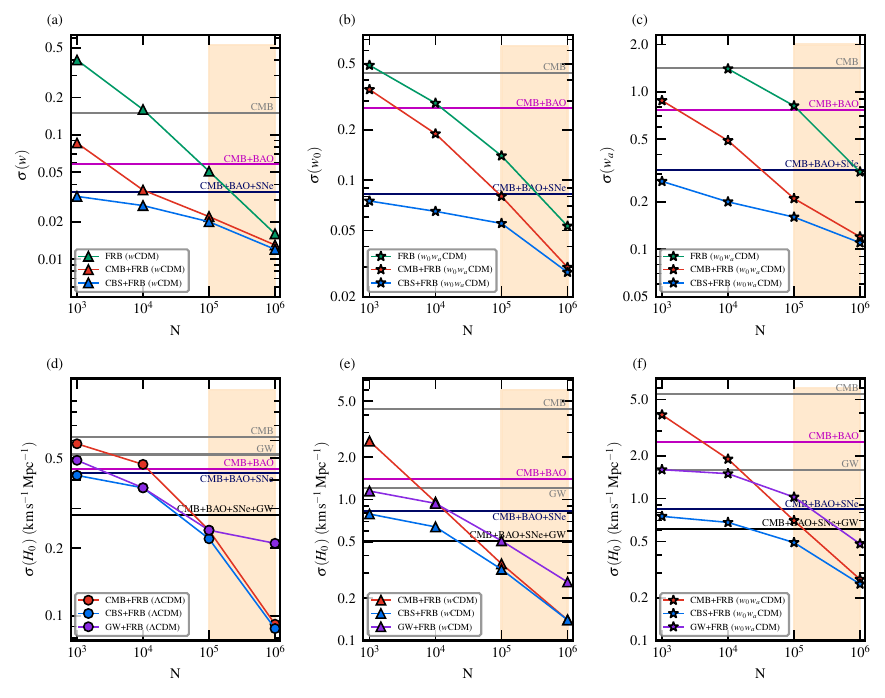}
\caption{The 1$\sigma$ absolute errors of dark-energy EoS parameters ($w$, $w_{0}$ and $w_{a}$) [(a)--(c)], and the Hubble constant $H_{0}$ [(d)--(f)] derived from $10^{3}$, $10^{4}$, $10^{5}$ and $10^{6}$ simulated FRB data (in logarithmic coordinates). The cyan lines represent the constraints from FRB data alone, while the red, blue and purple lines represent the results from the data combinations CMB+FRB, CBS+FRB and GW+FRB, respectively. The circle, triangle and star symbols denote the results from FRB data simulated in the $\Lambda$CDM, $w$CDM and $w_{0}w_{a}$CDM models, respectively. The grey horizontal lines represent the uncertainties given by CMB or GW alone, while the magenta, dark blue and black ones denote the uncertainties given by the CMB+BAO, CBS and CMB+BAO+SNe+GW data, respectively. The bisque region corresponds to the detection event rate given by the SKA. Note that $H_0$ is in units of km s$^{-1}$ Mpc$^{-1}$.} \label{wh0}
\end{figure*}

In order to comprehensively discuss what role the numbers of FRB data could play in the constraints on cosmological parameters, we present the constraint errors of dark-energy EoS parameters (i.e., $w$, $w_0$ and $w_a$) for $10^3$, $10^4$, $10^5$ and $10^6$ FRB data in the upper panels of Figure~\ref{wh0}. We find that the errors of these constraints from the FRB data alone, are almost consistent with the $1/\sqrt{N}$ behavior, with $N$ the number of data. 

We present a summary of the capability of future FRB data in constraining the dark-energy EoS parameters, based on our results of the $w$CDM and $w_{0}w_{a}$CDM models shown in the upper panels of Figure~\ref{wh0}. Using localized FRBs alone to measure dark energy requires a minimum of $10^4$ to provide comparable constraints as those from CMB. However, even with this number of FRB data, the constraints are not yet precise. An effective approach is that using FRBs as a complementary cosmological probe to break the parameter degeneracy induced by CMB. For example, using $10^4$ FRB data is comparable to using BAO data in breaking the parameter degeneracy of CMB \cite{zhao2020cosmological}.

Another approach could be directly utilizing a larger FRB sample, such as the data expected to be accumulated during the SKA era. At that time, the FRB data alone have the potential to break the parameter degeneracy of CMB even more effectively than BAO+SNe. Furthermore, the SKA-era FRB data alone can provide constraints at least comparable to those from the CBS data, even without the help of CMB, and independently and precisely measure dark energy in the late universe. Overall, in the era of SKA, localized FRBs can be forged into one of the precise cosmological probes in exploring the late universe to study dark energy.

\subsection{The Hubble constant}

\begin{figure*}
\setlength{\abovecaptionskip}{-0.2cm}
\begin{center}
\hspace*{.1cm}
\includegraphics[width=0.4\linewidth,angle=0]{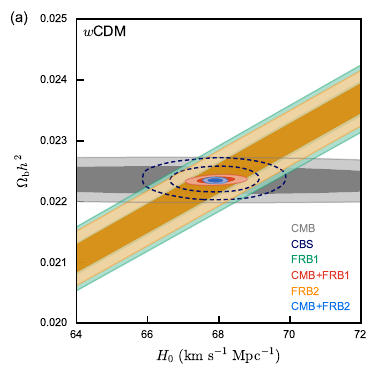}
\hspace*{.1cm}
\includegraphics[width=0.4\linewidth,angle=0]{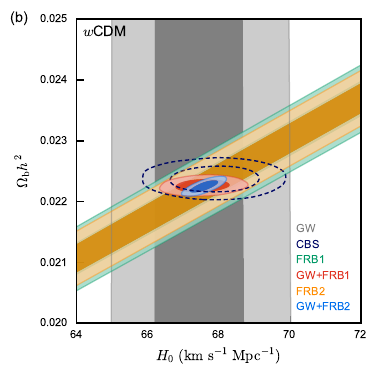}
\end{center}
\caption{Two-dimensional marginalized contours ($68.3\%$ and $95.4\%$ confidence level) in the $H_{0}$--$\Omega_{\rm b}h^{2}$ plane for the $w$CDM model by using the CMB, FRB1, CMB+FRB1, FRB2 and CMB+FRB2 data (a), and the GW, FRB1, GW+FRB1, FRB2 and GW+FRB2 data (b). For comparison, the dashed contours of CBS are added in both panels.} \label{Fig.h0obh2}
\end{figure*}

Then we report the measurement of the Hubble constant. Using FRB data alone can hardly constrain $H_0$ since $\rm{DM}_{\rm{IGM}} \propto \Omega_{\rm b}$$H_0$ (see Eq.~(\ref{eq3})), which results in the strong degeneracy between $H_0$ and $\Omega_{\rm b}h^{2}$. However, the early-universe measurement by CMB can effectively constrain $\Omega_{\rm b}h^{2}$. Therefore, combining the CMB and FRB data can break the degeneracy and obtain effective constraints on $H_{0}$.

For the $\Lambda$CDM model, the CMB data can precisely constrain $H_0$ with a 0.9$\%$ measurement, but the inclusion of the FRB2 data can improve this by nearly an order of magnitude to almost $0.1\%$. 
Even for the dynamical dark energy models, the combination of the CMB and FRB data can constrain the relative error of $H_{0}$ to less than $1\%$. Figure~\ref{Fig.h0obh2}(a) shows the contours of CMB, CBS, FRB1, CMB+FRB1, FRB2 and CMB+FRB2 in the $H_0$--$\Omega_{\rm b}h^{2}$ plane for the $w$CDM model, indicating that the degeneracy between $H_0$ and $\Omega_{\rm b}h^{2}$ inherent in FRB can be effectively broken by CMB+FRB1 and CMB+FRB2. Concretely, CMB+FRB1 gives the constraints of $\varepsilon(H_0) = 0.5\%$ and $1.0\%$ in the $w$CDM and $w_{0}w_{a}$CDM models, respectively. Compared to using the CMB data alone, this combination improves the constraints by about $93\%$ and $86\%$, respectively. CMB+FRB2 further improves these constraints, giving $\varepsilon(H_0) = 0.2\%$ and $0.4\%$ for the $w$CDM and $w_{0}w_{a}$CDM models, respectively, with improvements of about $94\%$ and $97\%$ compared to using the CMB data alone, respectively.

However, the joint analysis of CMB and FRB may pose challenges as it involves combining observations from both the early and the late universe, and the Hubble tension reflects the tension between the early and the late universe \citep{Verde:2019ivm}. As a result, it is not appropriate using CMB+FRB to study the Hubble tension. Thus, it is essential to investigate the estimation solely from the late universe, potentially providing a cross-check to the Hubble tension.

The GW standard siren detection can precisely constrain the Hubble constant, as the luminosity distance $d_{\rm{L}}$ to a GW source can be directly obtained from its waveform and then the Hubble constant through the relation of $d_{\rm{L}} \propto 1/H_0$. Therefore, combining the GW and FRB data can also give the effective constraints on $H_0$. Hence, we investigate the potential benefits of combining the GW standard siren observation with the ET and the FRB observation with the SKA in the coming decades. The constraint results from GW, FRB and GW+FRB are summarized in Table~\ref{tab:GW}.

The ET's GW standard siren data alone can give a $1.8\%$ constraint on $H_0$ in the $w$CDM model, which is better than the CMB measurement. However, combining the GW and FRB data can effectively constrain $H_0$ better than the CBS data, which is illustrated in Figure~\ref{Fig.h0obh2}(b). Compared to using the GW data alone, the inclusion of the FRB1 and FRB2 data can improve the constraints on $H_0$ by about $58\%$ and $78\%$, respectively. About other cosmological parameters, while the GW standard siren data can only give a $14\%$ constraint on $w$, the FRB2 data can give significantly better constraints (with a precision of $1.6\%$). The errors of constraints from FRB and GW+FRB are very similar. Hence, we do not discuss the dark-energy EoS constraints from GW+FRB.

We also present a summary of the capability of future localized FRBs in combination with GWs as an independent probe of low-redshift cosmic expansion, as shown in the bottom panels of Figure~\ref{wh0}. Combining $10^3$ or $10^4$ FRBs with GWs can provide $H_0$ constraints that are roughly comparable to those from CMB+BAO. Nevertheless, using $10^5$ SKA-era FRBs in combination with GWs can lead to even better results than current CBS. Moreover, by using $10^6$ FRBs in this combination, we can determine $H_0$ with a precision of less than $1\%$ in all the models, meeting the standards of precision cosmology. This indicates that the combination of future GW and FRB observations can serve as a powerful independent probe to study the late universe expansion history and the Hubble tension.

\subsection{Baryon density}

\begin{table*}[!htbp]
\caption{The relative errors (1$\sigma$) on the cosmological parameters related to baryon density ($\Omega_{\rm b}$, $\Omega_{\rm b}h$ and $\Omega_{\rm b}h^{2}$) in the $\Lambda$CDM model using the FRB1, $H_0$+FRB1, GW+FRB1, FRB2, $H_0$+FRB2 and GW+FRB2 data. Here $H_0$ denotes the local Hubble constant measurement of $H_{0}= 73.04 \pm 1.04 \rm{~km} \rm{~s}^{-1} \rm{Mpc}^{-1}$.}
\label{tab:baryon}
\setlength{\tabcolsep}{3mm}
\renewcommand{\arraystretch}{1.5}
\begin{center}{\centerline{
\begin{tabular}{m{1.5cm}<{\centering}m{1.5cm}<{\centering}m{1.5cm}<{\centering}m{1.5cm}<{\centering}m{1.5cm}<{\centering}m{1.5cm}<{\centering}m{1.5cm}<{\centering}m{1.5cm}<{\centering}}
\hline \hline
Model & Error 
& FRB1 & $H_0$+FRB1 & GW+FRB1 
& FRB2 & $H_0$+FRB2 & GW+FRB2 \\
\hline
\multirow{6}{*}{$\Lambda$CDM}
& $\sigma(\Omega_{\rm b})/10^{-3}$  
& 4.00  & 0.66 & 0.26
& 2.10  & 0.64 & 0.16 \\
& $\sigma(\Omega_{\rm b}h)/10^{-3}$  
& 0.086  & 0.086 & 0.071
& 0.027  & 0.027 & 0.026 \\
& $\sigma(\Omega_{\rm b}h^{2})/10^{-3}$  
& 4.10  & 0.35 & 0.070
& 4.05  & 0.34 & 0.066 \\
& $\varepsilon(\Omega_{\rm b})$  
& $8.2\%$  & $1.5\%$ & $0.5\%$
& $4.3\%$  & $1.4\%$ & $0.3\%$ \\
& $\varepsilon(\Omega_{\rm b}h)$  
& $0.3\%$  & $0.3\%$ & $0.2\%$
& $0.1\%$  & $0.1\%$ & $0.1\%$ \\
& $\varepsilon(\Omega_{\rm b}h^{2})$  
& $17.2\%$ & $1.4\%$ & $0.3\%$
& $17.0\%$ & $1.4\%$ & $0.3\%$ \\
\hline \hline
\end{tabular}}}
\end{center}
\end{table*}

\begin{figure*}[htb]
\includegraphics[width=0.32\linewidth,angle=0]{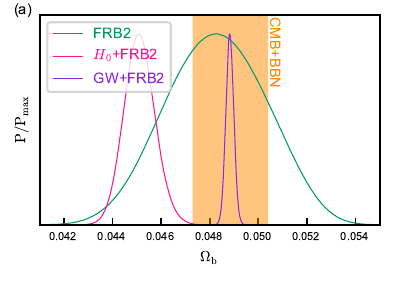}
\includegraphics[width=0.32\linewidth,angle=0]{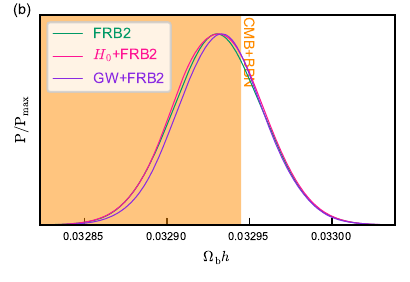}
\includegraphics[width=0.32\linewidth,angle=0]{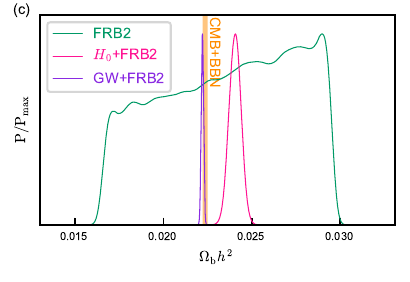}
\caption{One-dimensional marginalized probability distributions of $\Omega_{\rm b}$ (a), $\Omega_{\rm b}h$ (b) and $\Omega_{\rm b}h^{2}$ (c) by using the FRB2, $H_0$+FRB2 and GW+FRB2 data. The orange vertical regions denote the constraint results from CMB+BBN \citep{Pitrou:2018cgg}.}
\label{baryon}
\end{figure*}

Finally, we report the results on the cosmic baryon density estimated from the SKA-era FRBs. The parameters $\Omega_{\rm b}$ and $\Omega_{\rm b}h$ have already been constrained in literature, such as \citet{Yang:2022ftm} and \citet{macquart2020census}, respectively. Thus, in addition to considering the case of free parameters of $\Omega_m$, $H_0$ and $\Omega_{\rm b}h^{2}$ as the previous subsections, we also considered these two cases of free parameters, i.e. (i) $\Omega_m$, $H_0$ and $\Omega_{\rm b}$; (ii) $\Omega_m$ and $\Omega_{\rm b}h$. We use FRB1 and FRB2 to determine the credible intervals of $\Omega_{\rm b}$, $\Omega_{\rm b}h$ and $\Omega_{\rm b}h^{2}$ in the $\Lambda$CDM model. For the priors of $\Omega_{\rm b}$ and $\Omega_{\rm b}h$, we use uniform distributions within ranges of [0, 0.11] and [0, 0.07], respectively. The priors for the other parameters are the same as previous. In addition, we also consider combining the FRB data with the local $H_0$ measurement of $H_{0}= 73.04 \pm 1.04 \rm{~km} \rm{~s}^{-1} \rm{Mpc}^{-1}$ \citep{Riess:2021jrx}, i.e., $H_0$+FRB, to improve the constraints. The constraint results from using FRB, $H_0$+FRB and GW+FRB are summarized in Table~\ref{tab:baryon}.

We find that using only the FRB data provides very tight constraints on $\Omega_{\rm b}h$, which is better than those on $\Omega_{\rm b}$ and $\Omega_{\rm b}h^{2}$. This is because FRBs can directly constrain $\Omega_{\rm b}h$ while measuring the others may lead to related degeneracies
(see Eq.~(\ref{eq3})). For example, FRB1 gives a $0.3\%$ result of $\sigma(\Omega_{\rm b}h)=0.000086$, while $\Omega_{\rm b}$ and $\Omega_{\rm b}h^{2}$ are poorly constrained to $8.2\%$ and $17.2\%$, respectively.


On combining the FRB data with the late-universe $H_0$ measurement, the constraints on $\Omega_{\rm b}$ and $\Omega_{\rm b}h^{2}$ can be well improved when compared to those from the FRB data alone. For instance, $H_0$+FRB1 gives a constraint result of $\sigma(\Omega_{\rm b}h^{2})=0.00035$, which is a $92\%$ improvement compared to FRB1 alone. For $\Omega_{\rm b}h$, the constraints from $H_0$+FRB are not improved, since $H_0$ is included in $\Omega_{\rm b}h$, and therefore cannot be set as an independent free parameter. Nevertheless, the results on $\Omega_{\rm b}h$ from $H_0$+FRB are still better than those on $\Omega_{\rm b}$ and $\Omega_{\rm b}h^{2}$. In Figure~\ref{baryon}, we plot the one-dimensional posterior PDFs of $\Omega_{\rm b}$, $\Omega_{\rm b}h$ and $\Omega_{\rm b}h^{2}$ using the FRB2, $H_0$+FRB2 and GW+FRB2 data. We find that, the inclusion of the low-redshift $H_0$ measurement does not affect the best-fit value of $\Omega_{\rm b}h$, but it significantly impacts $\Omega_{\rm b}$, leading to a discrepancy ($\sim 1.5\sigma$) caused by the parameter degeneracy between $H_0$ and $\Omega_{\rm b}$.
Due to the Hubble tension, the actual value of $H_0$ is ambiguous, indicating that the Hubble tension may affect the local baryon census by FRB when discussing baryon density $\Omega_{\rm b}$ directly, but discussing $\Omega_{\rm b}h$ can effectively avoid the bias introduced by the Hubble tension.


On combining the FRB data with the GW standard siren data, we find that this combination can give significantly better constraint results on all cases of $\Omega_{\rm b}$, $\Omega_{\rm b}h$ and $\Omega_{\rm b}h^{2}$ than combining the late-universe $H_0$ measurement. In fact, the constraints on $\Omega_{\rm b}h^{2}$ are even better than that from CMB+BBN observations. For example, GW+FRB1 gives $\sigma(\Omega_{\rm b}h^{2})=0.000070$, which is twice as precise as $\sigma(\Omega_{\rm b}h^{2})=0.00014$ from CMB+BBN. This effect can be also seen in Figure~\ref{baryon}(c). The GW data can effectively constrain the Hubble constant, and its inclusion can significantly improve the FRB constraint on $\Omega_{\rm b}h^{2}$ due to the strong degeneracy between $H_0$ and $\Omega_{\rm b}h^{2}$. This is similar to the method of combining CMB+FRB to constrain $H_0$. So, this joint approach may offer a very precise late-universe probe of the cosmic baryon density, independent of the early-universe CMB+BBN data. Note that the fiducial values of simulated GW data are consistent with the CBS data, so the constraints from GW+FRB are consistent with the CMB+BBN results but in tension with the $H_0$+FRB results in the $\Omega_{\rm b}$ and $\Omega_{\rm b}h^{2}$ estimations.


By comparing the results of individual and joint analyses on different baryon density parameters, we conclude that using FRBs alone to constrain $\Omega_{\rm b}h$ seems appropriate for a local baryon census. There are two reasons: (i) discussing $\Omega_{\rm b}h$ could be precise, as using FRB alone can provide a very precise constraint on $\Omega_{\rm b}h$ without the need to combine other data; (ii) discussing $\Omega_{\rm b}h$ could be accurate, effectively avoiding the bias introduced by the Hubble tension.

\section{CONCLUSION}\label{sec4}
Despite the current cosmology keeps in good concordance, a series of puzzles remain, including dark energy, the Hubble tension and the missing baryon problem. This work investigates how upcoming FRB observation with the SKA can help solve these problems. We consider two scenarios, one with an normal expected detection rate of $\sim 10^4$ FRBs per year and another with an optimistic rate of $\sim 10^5$ FRBs per year, and construct mock catalogues of $10^5$ and $10^6$ localized FRBs, respectively, over a 10-year operation time of the SKA. We use MCMC techniques to forecast constraints for cosmological parameters in three typical dark energy models, i.e., the $\Lambda$CDM, $w$CDM and $w_{0}w_{a}$CDM models. We have the following main findings.


\begin{itemize}

\item The issue of dark energy. We find that, solely using $10^6$ FRB data can give very tight constraints on dark-energy EoS parameters, with $\sigma(w)=0.016$ in the $w$CDM model, and $\sigma(w_0)=0.053$ and $\sigma(w_a)=0.31$ in the $w_{0}w_{a}$CDM model. These results are about $3\%$--$54\%$ better than those from the current mainstream CBS data. Although the use of $10^5$ FRBs could not achieve this precision, it is more effective in breaking the inherent parameter degeneracies in CMB than BAO+SNe. The joint CMB+FRB data in the $w$CDM model gives $\sigma(w)=0.022$, and in the $w_{0}w_{a}$CDM model, it gives $\sigma(w_0)=0.080$ and $\sigma(w_a)=0.21$. These results are about over $80\%$ and $30\%$ better than the results by CMB and CBS for both models, respectively. We conclude that during the SKA era, localized FRBs can be developed as a precise cosmological probe for exploring the late universe. This approach has the potential to provide a way to explore dark energy using only one cosmological probe.
 

\item The issue of the Hubble constant. We combine the SKA-era FRB data with the CMB or simulated GW standard siren data from ET's 10-year observation, and find that the combinations can effectively constrain the Hubble constant. This is because CMB and GW standard siren can precisely constrain $\Omega_{\rm b}h^{2}$ and $H_0$, respectively, making the inherent parameter degeneracies in the FRB data be broken. When combined with the GW data, we find that using $10^5$ FRB data can give more precise measurements than CBS, and using $10^6$ FRB data provides constraint precision of less than $1\%$ in all the models, meeting the standard of precision cosmology. We conclude that both the joint CMB+FRB and GW+FRB results provide powerful methods to study the expansion history of the universe, and particularly the GW+FRB results, being a late-universe measurement, can serve as an independent cross-check to help study the Hubble tension.

\item The issue of baryon density. We compare the constraints on baryon density parameters $\Omega_{\rm b}$, $\Omega_{\rm b}h$ and $\Omega_{\rm b}h^{2}$ under $\Lambda$CDM, and find that using the FRB data alone can give very tight constraints on $\Omega_{\rm b}h$. In order to improve the constraints on $\Omega_{\rm b}$ and  $\Omega_{\rm b}h^2$ and simultaneously obtain a late-universe result, we can combine the local $H_0$ measurement and FRB data. However, when applying the local $H_0$ prior, the Hubble tension may affect the local baryon census by FRB when discussing  $\Omega_{\rm b}$ directly, but discussing $\Omega_{\rm b}h$ can avoid this bias. Furthermore, combining the GW and FRB data can significantly improve the constraints on $\Omega_{\rm b}h^2$, providing a very precise late-universe probe that is independent of CMB+BBN data. We conclude that using FRBs alone to constrain $\Omega_{\rm b}h$ can precisely and accurately measure baryon density. 



\end{itemize}

Dark energy cosmology serves as the key project for the SKA. After the completion and deployment of SKA in the 2030s, an extensive dataset of FRB measurements with redshift information will become available. This will provide us with the opportunity to utilize FRB observations to precisely measure the evolutionary history of the universe. Moreover, other wide-field telescopes such as the Canadian Hydrogen Observatory and Radio transient Detector (CHORD) \citep{van2019}, Hydrogen Intensity and Real-time Analysis eXperiment (HIRAX) \citep{new2016hirax}, and DSA-2000 \citep{hal2019} are also anticipated to contribute to the field by delivering additional samples of localized FRBs. The landscape of FRB cosmology will be very bright and many exciting science returns can be expected in the coming years.

\acknowledgments
We are very grateful to Peng-Ju Wu, Shang-Jie Jin, Ling-Feng Wang, Jing-Zhao Qi and Chen-Hui Niu for helpful discussions. This work was supported by the National SKA Program of China (Grants Nos. 2022SKA0110200 and 2022SKA0110203), the National Natural Science Foundation of China (Grants Nos. 11975072, 11835009, 11875102, and 11988101), and the National 111 Project of China (Grant No. B16009).

\bibliography{frbska}

\end{document}